\documentclass[10pt, a4paper]{article}
\usepackage[margin=1.1in]{geometry}

\usepackage{natbib}
\usepackage{amssymb}
\usepackage{amsmath}
\usepackage{amsthm}
\usepackage[normalem]{ulem}
\usepackage{graphicx}
\usepackage{caption}
\usepackage{subcaption}
\usepackage{float}   
\usepackage{placeins}  
\usepackage[colorinlistoftodos, textwidth=4cm, shadow]{todonotes}

\newtheorem{problem}{Problem}

\usepackage[linesnumbered,ruled,vlined]{algorithm2e}
\usepackage{mathtools}

\usepackage[toc,page,header]{appendix}
\usepackage{minitoc}
\usepackage{hyperref}
\hypersetup{
    colorlinks=true,
    linkcolor=black,   
    citecolor=black,   
    urlcolor=blue      
}

\title{GOGH: Correlation-Guided Orchestration of GPUs in Heterogeneous Clusters}
\author{
  Ahmad Raeisi, Mahdi Dolati, Sina Darabi, Sadegh Talebi, Patrick Eugster, and Ahmad Khonsari\footnote{
  Affiliations: 
  University of Tehran; 
  Sharif University of Technology; 
  Università della Svizzera italiana (USI); 
  University of Copenhagen.\\
  Emails: {\tt\small ahmad.raeisi@ut.ac.ir, mdolati@sharif.edu, sina.darabi@usi.ch, m.shahi@ku.dk, patrick.eugster@usi.ch, khonsari@ut.ac.ir}.
  }
}

\date{\today}

\begin{document}
\maketitle

\begin{abstract}
    The growing demand for computational resources in machine learning has made efficient resource allocation a critical challenge, especially in heterogeneous hardware clusters where devices vary in capability, age, and energy efficiency. Upgrading to the latest hardware is often infeasible, making sustainable use of existing, mixed-generation resources essential. In this paper, we propose a learning-based architecture for managing machine learning workloads in heterogeneous clusters. The system operates online, allocating resources to incoming training or inference requests while minimizing energy consumption and meeting performance requirements. It uses two neural networks: the first provides initial estimates of how well a new model will utilize different hardware types and how it will affect co-located models. An optimizer then allocates resources based on these estimates. After deployment, the system monitors real performance and uses this data to refine its predictions via a second neural network. This updated model improves estimates not only for the current hardware but also for hardware not initially allocated and for co-location scenarios not yet observed. The result is an adaptive, iterative approach that learns over time to make more effective resource allocation decisions in heterogeneous deep learning clusters.
\end{abstract}

\section{Introduction}
Recent breakthroughs in machine learning have driven significant progress in domains such as image recognition, natural language processing, and scientific computing~\cite{wang2024yolov10,liu2024vmamba}. These advancements have been fueled by increasingly large models and datasets, resulting in substantial computational demands. To meet these demands, there has been a surge in the design and deployment of specialized hardware accelerators~\cite{hanindhito2024accelerating}. However, due to cost and operational constraints, upgrading all infrastructure to the latest hardware is often impractical. As a result, sustainable machine learning infrastructure must be designed to operate across heterogeneous environments~\cite{um2024metis}, incorporating hardware of varying generations. Effectively managing such clusters requires systems that can adapt to differences in performance and capabilities, ensuring efficient utilization of available resources~\cite{Mei_ASPLOS_2025}.

The management of heterogeneous hardware for machine learning tasks presents a significant challenge within the field. A primary difficulty stems from the limited understanding of how effectively various machine learning jobs can leverage distinct hardware architectures~\cite{weng2022mlaas}. Machine learning models display considerable diversity in both structural design and computational demands, leading to potential conflicts when multiple models are deployed contemporaneously on shared hardware resources, thereby diminishing overall system efficiency. Predicting the performance of individual models on specific hardware configurations and understanding the interactions among co-located models pose further complexities~\cite{yeung2020towards}. Nonetheless, there exist discernible patterns and commonalities among different models and hardware platforms that offer solutions to these issues. Leveraging historical performance data can provide insightful forecasts regarding the behavior of novel models on varied hardware, as well as their potential interactions when resources are concurrently utilized~\cite{xiao2018gandiva}.

In this paper, we propose the correlation-Guided Orchestration of GPUs in Heterogeneous clusters (GOGH), a framework for assigning incoming deep learning jobs to heterogeneous GPUs in a deep learning cluster. GOGH leverages historical execution data to predict the throughput of different jobs on various GPUs, enabling more informed scheduling decisions. GOGH exploits two key types of correlation: (i) the similarity between different jobs, and (ii) the variation in throughput of the same job across different GPU types. GOGH consists of three core modules that operate in an online and iterative fashion. The first module provides an initial throughput estimate for newly arrived jobs using a nearest-neighbor approach guided by prior observations. The second module utilizes these estimates to allocate jobs to GPUs while optimizing for energy efficiency and ensuring minimum throughput guarantees for all jobs. The third module refines the throughput estimates based on observed performance after job execution, thereby improving the accuracy of future GPU allocations. To implement the first and third modules, we employ neural networks and investigate the impact of model choice—specifically, LSTM and Transformer architectures—on estimation accuracy. For the second module, we formulate the GPU allocation task as an integer linear programming (ILP) problem and utilize standard solvers to compute optimal assignments. Experimental results using a publicly available dataset~\cite{narayanan2020heterogeneity} demonstrate the effectiveness of GOGH in improving both scheduling efficiency and prediction accuracy. The main contributions of this work are as follows:

\begin{itemize}
    \item We propose GOGH, a novel framework for orchestrating deep learning workloads across heterogeneous GPU clusters. GOGH leverages historical performance data to guide scheduling decisions, accounting for both inter-job and inter-GPU correlations.
    \item We formulate the GPU allocation problem as an integer linear programming (ILP) model that jointly considers throughput maximization, energy efficiency, and minimum performance guarantees for jobs.
    \item We explore different neural architectures for the throughput estimation module, and analyze their impact on prediction accuracy and scheduling performance.
    \item We validate the effectiveness of GOGH using a publicly available dataset of deep learning jobs and heterogeneous GPU performance logs. Results demonstrate significant improvements in both scheduling efficiency and throughput estimation accuracy compared to baseline methods.
\end{itemize}

\section{GOGH Design}
\begin{figure}
    \centering
    \includegraphics[width=0.8\linewidth]{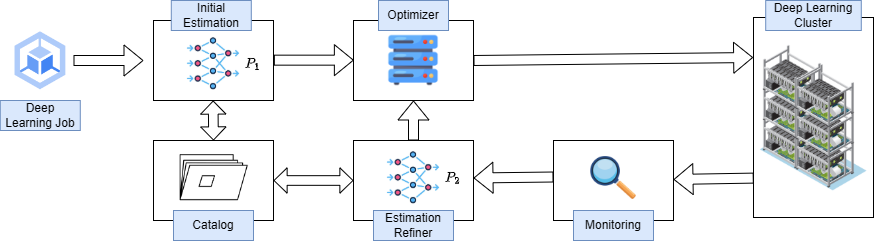}
    \caption{The architecture of GOGH.}
    \label{fig:arch}
\end{figure}

In this section, we introduce GOGH, a framework for online GPU allocation in a heterogeneous deep learning cluster. Machine learning jobs arrive over time, and the system must dynamically decide which GPUs to assign to each job. We assume no prior knowledge of how each job will perform on different GPU types. Figure~\ref{fig:arch} illustrates the high-level architecture of GOGH. 

\subsection{High-Level Description}
GOGH employs two neural networks, $P_{1}$ and $P_{2}$, to guide its decision-making process. The first network, $P_{1}$ , is responsible for generating initial estimates of a new job’s throughput on each type of GPU, as well as estimating the impact of the new job on the throughput of other jobs already running on the same GPU. To compute these estimates, $P_{1}$ relies on historical data from previously executed jobs in the cluster. This data is maintained by a module called Catalog, which stores the measured or estimated throughput of each job on each GPU type. These values are continuously updated to reflect the most recent and accurate information. When a new job arrives, $P_{1}$ retrieves the most similar previously seen job from the Catalog—based on feature similarity—and uses the corresponding throughput records to estimate the new job’s performance. Using these estimates, a module called the Optimizer selects a GPU allocation that satisfies the job’s throughput requirements while minimizing overall energy consumption.

Once a job is assigned to a set of GPUs and begins execution, a monitoring module measures its actual throughput on each GPU. This real-world performance data is then fed to the second neural network, $P_{2}$, which also has access to the latest throughput estimates stored in the Catalog. Specifically, $P_{1}$ writes its initial estimates to the Catalog, and $P_{2}$ updates these values whenever new monitoring data becomes available, recalculating estimates accordingly. This feedback loop enables the system to continuously learn and improve its resource allocation strategy over time. Importantly, the new monitoring data is also used to refine estimates of how the same job would perform on GPU types it was not allocated to. In addition, $P_{2}$ updates throughput estimates for co-location scenarios—i.e., how the job performs when sharing a GPU with other jobs in the cluster. Through this continuous learning process, the system develops a more comprehensive and generalizable understanding of job–GPU compatibility, enabling more effective and informed future allocations.

\begin{table}[t]
    \centering
    \caption{Important Notations.}
    \label{tab:notation}
    \fontsize{8}{9}\selectfont
    \begin{tabular}{|c|p{12cm}|} \hline
        \textbf{Symbol} & \textbf{Definition} \\ \hline\hline
        $\mathcal{S}$ & Set of all servers \\ \hline
        $\mathcal{A}$ & Set of all accelerator types \\ \hline
        $\mathcal{J}$ & Set of all jobs \\ \hline
        $\mathcal{C}$ & Set of job combinations \\ \hline
        $\mathcal{C}_{j}$ & Set of job combinations containing job $j$ \\ \hline
        $\vert c\vert$ & Number of jobs in combination $c\in\mathcal{C}$ \\ \hline
        $T_{a,j}^{c}$ & Throughput of job $j$ on accelerator of type $a$ when combination $c$ is assigned \\ \hline
        $\overline{T}_{j}$ & Minimum required throughput of job $j$ \\ \hline
        $\theta_{a}$ & Job capacity of accelerator of type $a$ \\ \hline 
        $D_{j}$ & Distributability of job $j$ \\ \hline 
        $x_{a,s}^{c}$ & Placement of combination $c$ on accelerator of type $a$ in server $s$ \\ \hline
    \end{tabular}
\end{table}

\subsection{Notation}
To formally define the problem, let $\mathcal{S}$ denote the set of all servers in the cluster, and let $\mathcal{A}$ be the set of all accelerator types (i.e., different GPU types). Let $\mathcal{J}$ represent the set of all jobs active in the cluster at the time the optimizer is executed. Let $\Psi_{j}$ denote the attribute vector that characterizes machine learning job $j$—such as model architecture (e.g., ResNet) and batch size. Define $\mathcal{C}\subseteq \mathcal{P}(\mathcal{J})$ as the set of valid job combinations, where $\mathcal{P}(.)$ denotes the powerset function. Since most accelerators support only one or two co-located jobs, we restrict $\mathcal{C}$ to subsets of $\mathcal{J}$ of size one or two, without loss of generality. Let $\theta_{a}$ denote the capacity of an accelerator of type $a\in\mathcal{A}$, and use $\vert c\vert$ to denote the number of jobs in a combination $c\in\mathcal{C}$. Define $\mathcal{C}_{j}\subseteq\mathcal{C}$ as the set of job combinations that include job $j\in\mathcal{J}$. Let $T_{a,j}^{c}$ represent the measured throughput of job $j$ on an accelerator of type $a$ when the job combination $c$ is assigned to it; this serves as a performance measure. We denote by $\widetilde{T}{a,j}^{i,c}$ the $i$-th estimate of $T{a,j}^{c}$, obtained using the neural networks $P_1$ and $P_2$. Specifically, $P_1$ always provides the initial estimate $\widetilde{T}{a,j}^{0,c}$, while $P_2$ computes $\widetilde{T}{a,j}^{i,c}$ based on the measurement results from round $i$. A summary of the key notations is provided in Table~\ref{tab:notation}.

\subsection{Initial Throughput Estimation}
Upon the arrival of a new job $j_1$, GOGH selects a similar existing job $j_2$ from the Catalog by comparing their attribute vectors $\Psi_{j_1}$ and $\Psi_{j_2}$. GOGH then uses neural network $P_1$ to estimate the throughput of $j_1$ on each GPU type in two scenarios: (i) when $j_1$ runs alone on the GPU, i.e., $\widetilde{T}{a,j_1}^{0,c}$ for $c = {j_1}$, and (ii) when $j_1$ shares the GPU with another job, i.e., $\widetilde{T}{a,j_1}^{0,c}$ for $\lvert c \rvert = 2$ and $j_1 \in c$. The inputs and outputs of $P_1$ are described formally below:
\begin{align}
    (
        \Psi_{j_{2}},
        \Psi_{j_{3}},
        a,
        T_{a,j_{2}}^{\{j_{2},j_{3}\}},
        T_{a,j_{3}}^{\{j_{2},j_{3}\}},
        \Psi_{j_{1}}
    )
    \rightarrow 
    (
        \widetilde{T}_{a,j_{1}}^{0,\{j_{1},j_{3}\}},
        \widetilde{T}_{a,j_{3}}^{0,\{j_{1},j_{3}\}}
    ),
    \qquad
    \forall j_{3}\in\mathcal{J}\cup\{j_{0}\}
\end{align}
The underlying assumption is as follows: if jobs $j_2$ and $j_3$ co-execute on a GPU of type $a$, their throughputs are $T_{a,j_2}^{c}$ and $T_{a,j_3}^{c}$, respectively. We know the values of $T_{a,j_2}^{c}$ and $T_{a,j_3}^{c}$ from historical records stored in the Catalog. Since $j_2$ is similar to the incoming job $j_1$—as indicated by the similarity of $\Psi_{j_1}$ and $\Psi_{j_2}$—this relationship enables the estimation of the throughput of $j_1$ when it is colocated with $j_3$ on a GPU of type $a$. To handle cases where $j_2$ was the sole occupant of the GPU, we introduce a synthetic job $j_0$ with feature vector $\Psi_{j_0} = \mathbf{0}$ and throughput $T_{a,j_0}^{c} = 0$, representing an empty slot. In Subsection~\ref{ss_refine}, we detail how these initial estimates $\widetilde{T}_{a,j}^{0,c}$ are refined into $\widetilde{T}_{a,j}^{1,c}$ using runtime measurements collected from the system.

% Given a hypothetical modification to the attributes of job $j_1$, denoted by $\widetilde{\Psi}_{j_1}$, we aim to estimate the resulting throughput for the co-execution of the modified job with $j_2$ on the same accelerator.
\begin{problem}[GPU allocation problem]\label{problem_1}
\fontsize{8}{7}\selectfont
\begin{subequations}
  \begin{flalign}
    \min\limits_{x_{a,s}^{c}}\ & \sum_{a\in\mathcal{A}} \gamma_{a}\Big(\sum_{c\in\mathcal{C}_{j}}T_{a,j}^{c} \, x_{a,s}^{c}\Big) \label{eq_obj} &\\
    \text{s.t. } & \sum_{s\in\mathcal{S}}\sum_{a\in\mathcal{A}}\sum_{c\in\mathcal{C}_{j}} x_{a,s}^{c} \ge 1, & j\in\mathcal{J} \label{eq_c1}\\
    & \sum_{s\in\mathcal{S}}\sum_{a\in\mathcal{A}}\sum_{c\in\mathcal{C}_{j}} x_{a,s}^{c} \le D_{j}, & j\in\mathcal{J} \label{eq_c2}\\
    & \sum_{c\in\mathcal{C}} |c|\, x_{a,s}^{c} \le \theta_{a}, & s\in\mathcal{S},\, a\in\mathcal{A} \label{eq_c3}\\
    & \overline{T}_{j} \le \sum_{a\in\mathcal{A}}\sum_{c\in\mathcal{C}_{j}} T_{a,j}^{c}\, x_{a,s}^{c}, & j\in\mathcal{J} \label{eq_c4}\\
    & \sum_{c\in\mathcal{C}} x_{a,s}^{c} \le 1, & a\in\mathcal{A},\, s\in\mathcal{S} \label{eq_c5}\\
    & x_{a,s}^{c} \in \{0,1\} &
  \end{flalign}
\end{subequations}
\end{problem}

\subsection{Optimizer}
Once the throughput of each job on each GPU type is obtained—either via direct measurement or estimation—it becomes possible to allocate GPUs to jobs such that each job receives at least its minimum required throughput, denoted by $\overline{T}_{j}$, while minimizing overall power consumption and satisfying the capacity constraints of each GPU. To achieve this, we formulate an integer linear program (ILP), which we solve using a standard off-the-shelf optimization solver. In the ILP formulation, we introduce a binary decision variable $x_{a,s}^{c}$ that indicates whether the job combination $c$ is assigned to an accelerator of type $a$ on server $s$. The complete formulation is presented in Problem~\ref{problem_1}. The objective function, shown in~\eqref{eq_obj}, minimizes the total power consumption, where $\gamma_{a}(x)$ denotes the power usage of accelerator type $a$ under load $x$. The function $\gamma_{a}(\cdot)$ can be obtained through profiling methods such as those described in~\cite{10327871}. The constraints are defined as follows: Constraint~\eqref{eq_c1} ensures that each job is assigned to at least one accelerator. Constraint~\eqref{eq_c2} limits the number of accelerators assigned to job $j$ to at most $D_j$. Constraint~\eqref{eq_c3} enforces the capacity constraints of accelerators. Constraint~\eqref{eq_c4} guarantees that each job achieves at least its required throughput $\overline{T}_j$. Constraint~\eqref{eq_c5} ensures that each accelerator is assigned to at most one job combination. Although it is possible to linearize $\gamma_a(x)$ using piecewise approximations and leverage the fact that all constraints in Problem~\ref{problem_1} are either packing or covering constraints to develop more efficient algorithms, we leave this as an avenue for future work. In this study, we rely on a general-purpose solver to obtain high-quality solutions to Problem~\ref{problem_1}.

\subsection{Estimation Refinement} \label{ss_refine}
After the GPUs are allocated to jobs, GOGH continuously monitors the actual throughput of each job on each GPU type. Once these measurements become available, they are used to refine the throughput estimates. Specifically, GOGH leverages the correlation between a job’s throughput on different GPU types to improve the accuracy of our predictions. Suppose the estimated throughput of job $j_1$ in combination $c$ on GPU type $a_1$ is $\widetilde{T}{a_1,j_1}^{i-1,c}$, and the observed (measured) throughput is $T_{a_1,j_1}^{c}$. This observation enables us to refine the estimate of the same job in the same combination on a different GPU type $a_2 \ne a_1$, i.e., to improve $\widetilde{T}_{a_2,j_1}^{i-1,c}$. Neural network $P_2$ performs this refinement using the available information—specifically, the estimated and observed throughputs on $a_1$ for both $j_1$ and its colocated job $j_2$ (i.e., $\widetilde{T}{a_1,j_1}^{i-1,c}$, $\widetilde{T}_{a_1,j_2}^{i-1,c}$, $T_{a_1,j_1}^{c}$, and $T_{a_1,j_2}^{c}$)—to update the current estimates on GPU type $a_2$, namely $\widetilde{T}_{a_2,j_1}^{i-1,c}$ and $\widetilde{T}_{a_2,j_2}^{i-1,c}$. The inputs and outputs of $P_2$ are described formally below:
\begin{align}
    (
        \Psi_{j_{1}},
        \Psi_{j_{2}},
        a_{1},
        a_{2},
        &\widetilde{T}_{a_{1},j_{1}}^{i-1,c},
        \widetilde{T}_{a_{1},j_{2}}^{i-1,c}, \nonumber \\
        &T_{a_{1},j_{1}}^{c},
        T_{a_{1},j_{2}}^{c},
        \widetilde{T}_{a_{2},j_{1}}^{i-1,c},
        \widetilde{T}_{a_{2},j_{2}}^{i-1,c}
    )
    \rightarrow 
    (
        \widetilde{T}_{a_{2},j_{1}}^{i,c},
        \widetilde{T}_{a_{2},j_{2}}^{i,c}
    ),
    \qquad
    \forall j_{2}\in\mathcal{J}\cup \{j_{0}\}
\end{align}
Each new measurement of job $j$ in combination $c$ on a GPU type $a' \ne a$ provides an additional data point for refining the estimate on GPU type $a$. To track these refinements, we define the set $\mathcal{T}{a,j}^{c}$, which stores all updated values for $\widetilde{T}_{a,j}^{i-1,c}$ derived from observations on other GPU types:
\begin{equation}
    \mathcal{T}_{a,j}^{c} = \bigcup_{i} \widetilde{T}_{a,j}^{i,c}, 
\end{equation}
The final refined estimate $\widetilde{T}_{a,j}^{i,c}$ is then computed as the average of the values stored in $\mathcal{T}_{a,j}^{c}$.

\section{Evaluation}
In this section we present the evaluation of GOGH. 

\subsection{Setup}

% using Jupyter Notebook as the interactive development environment. All experiments were executed within a Python virtual environment to ensure dependency isolation and reproducibility. 

% , and NumPy 1.22.3 was used for numerical computations. All training and evaluation procedures were executed on the CPU, as GPU acceleration was not enabled in the TensorFlow configuration.

\noindent\textbf{System:} 
 The implementation was developed in Python 3.9.13 and the machine learning models were implemented using TensorFlow 2.11.0 with Keras as the high-level API. All measurements were taken under controlled conditions, ensuring consistent thermal and power states across experimental runs.

\noindent\textbf{Dataset:}
We use the dataset introduced in Gavel~\cite{narayanan2020heterogeneity}, a benchmark suite that provides detailed throughput measurements for a wide range of machine learning workloads running on heterogeneous GPU clusters. We consider the throughput as the primary metric. The data set in~\cite{narayanan2020heterogeneity} measures the throughput in iterations per second. For each job-accelerator combination, the dataset provides a solo throughput that shows the performance of a job running alone on a given accelerator and a co-located throughput that shows the throughputs of two simultaneously running jobs. The cluster consists of six types of GPU accelerators: \{k80, p100, v100, k80\_unconsolidated, p100\_unconsolidated, v100\_unconsolidated\}. These represent a realistic mix of legacy and modern hardware, with \texttt{*\_unconsolidated} variants capturing scenarios with fragmented or partially utilized resources. The workload set consists of deep learning models commonly used in production environments, including image classification (ResNet), natural language processing (Transformer, LM), and recommendation systems. Each workload is defined by a model type, a batch size, and a replication factor (fixed at 1 for this study). The detailed description of evaluated jobs is presented in Table~\ref{tab_workloads}. This selection covers a spectrum of workload profiles, from compute-intensive image models to memory-heavy recommendation systems and sequence-based NLP models.

% The dataset includes both solo execution metrics and co-location throughput values, enabling a comprehensive simulation of scheduling and allocation scenarios. 

% \subsubsection{Dataset and Cluster Configuration}

% \begin{itemize}
%     \item \textbf{ResNet-18, ResNet-50}: Batch sizes of 16, 32, 64, 128, 256
%     \item \textbf{Transformer}: Batch sizes of 16, 32, 128, 256
%     \item \textbf{Language Model (LM)}: Batch sizes of 5, 10, 20, 80
%     \item \textbf{Recommendation System}: Batch sizes of 512, 1024, 2048, 8192
% \end{itemize}

\begin{table}[t]
    \centering
    \caption{Workloads employed in the simulation.}
    \label{tab_workloads}
    \begin{tabular}{|c|p{2.7cm}|p{2.2cm}|p{2cm}|p{3cm}|}
        \hline
        \textbf{Application} & 
            ResNet-18/ResNet-50 &
            Transformer &
            Language Model (LM) &
            Recommendation System \\ \hline
        \textbf{Batch Size} & 
            $\{16, 32, 64, 128, 256\}$ &
            $\{16, 32, 128, 256\}$ &
            $\{5, 10, 20, 80\}$ &
            $\{512, 1024, 2048, 8192\}$ \\ \hline
    \end{tabular}
\end{table}

\noindent\textbf{Neural Network Implementation:}
To evaluate the effectiveness of GOGH, we implemented three neural network variants for both the initial prediction model (P1) and the refinement model (P2): 
\begin{itemize}
    \item Feedforward (FF),
    \item Recurrent Neural Network (RNN),
    \item Transformer.
\end{itemize}
All models were designed to maintain similar structural complexity, with comparable numbers of layers, hidden units, and training configurations. This uniformity ensures that observed performance differences stem from architectural characteristics rather than disparities in model capacity. Models were assessed using loss (mean squared error) and mean absolute error (MAE) across training, validation, and test sets.

% We quantitatively evaluate the performance of our two-stage learning framework by comparing three neural network architectures—Feedforward (FF), Recurrent Neural Network (RNN), and a customized Transformer—on both the initial estimation model (P1) and the refinement model (P2). 

% To explore the architectural impact on prediction quality, we designed and evaluated three neural network models for each stage: a fully connected feedforward network, a recurrent neural network (RNN), and a customized Transformer. Each model processes the same input structure but differs in how it captures relationships between job attributes and their interaction with accelerator types. This comparative study enables us to assess which architecture is best suited to the distinct learning goals of each stage.

% To ensure experimental reproducibility, random seed control was applied across all relevant libraries, including NumPy, Python’s built-in random module, and TensorFlow. This ensured consistent weight initialization, data shuffling, and model behavior across runs. 

% Matplotlib was used solely for visualizing results and was not involved in model computation or evaluation.

\subsection{Results}
In this subsection, we assess the effectiveness of GOGH in accurately predicting workload performance and improving resource allocation decisions in heterogeneous machine learning clusters. Specifically, we test whether the two-stage neural network architecture can (1) provide accurate initial throughput estimates for co-located jobs on a given accelerator, and (2) refine those estimates over time using real performance feedback to improve allocation decisions across different accelerator types. Additionally, we aim to compare the predictive capability of different neural network architectures to determine which model design yields better performance in this context. This helps us understand not only whether the system works, but which architectural choices contribute most effectively to throughput prediction accuracy.

% \subsection{Framework Workflow}
% The evaluation is centered on the effectiveness of the two neural network models in our framework:

% \begin{enumerate}[leftmargin=*]
%     \item \textbf{Initial Prediction Model}: Receives two job configurations and an accelerator type as input. It predicts the co-located throughput of both jobs when one of them is modified (e.g., batch size change), aiding in initial scheduling decisions.
    
%     \item \textbf{Refinement Model}: Continuously trained on real execution feedback post-deployment. This model generalizes throughput predictions to new accelerators and previously unseen co-location scenarios, improving decision quality over time.
% \end{enumerate}

% This two-stage approach mirrors the practical lifecycle of cluster scheduling: making informed initial allocations, then refining them as more data becomes available. The framework's adaptability is critical for sustained performance and energy efficiency in dynamic, heterogeneous environments.

% \textbf{Describe results qualitatively}

\begin{figure}[t]
    \centering
    \begin{subfigure}{0.48\linewidth}
        \centering
        \includegraphics[width=1\linewidth]{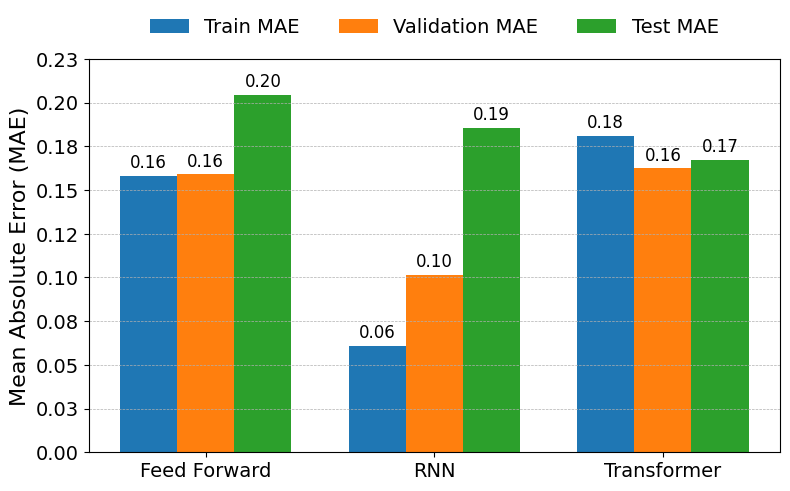}
    \caption{Initial NN (P1).}
    \label{fig:performance-p1}
    \end{subfigure} \hfill
    \begin{subfigure}{0.48\linewidth}
        \centering
        \includegraphics[width=1\linewidth]{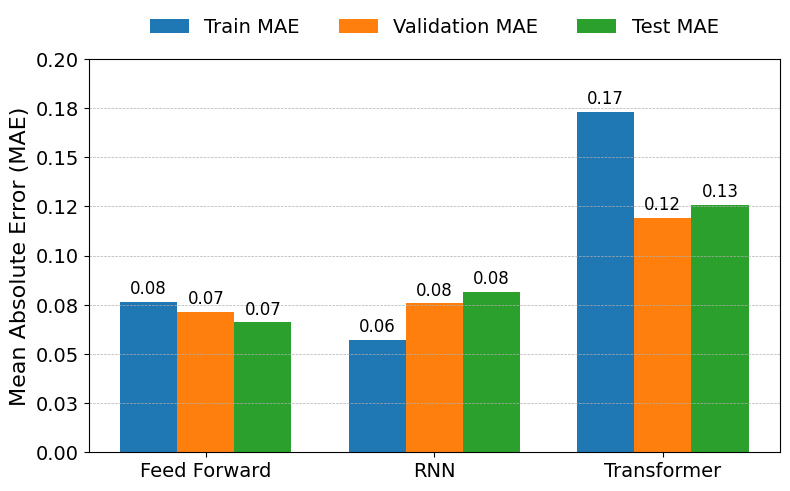}
        \caption{Refinement NN (P2).}
        \label{fig:performance-p2}
    \end{subfigure} 
    \caption{Mean Absolute Error (MAE) of different models at initial estimation and estimation refinement.}
    \label{fig_ut_batch}
\end{figure}

\noindent\textbf{Initial Estimation (P1):}  
The results for the performance metrics of the initial estimation across training, validation, and test sets for all three models are summarized in Figure~\ref{fig:performance-p1}. On the validation set, the RNN model outperformed the others, achieving a MAE of $0.1013$, which is $36.2\%$ lower than that of the Feedforward model (FF) ($0.1589$) and $37.6\%$ lower than that of the Transformer ($0.1624$). Similarly, in terms of training MAE, the RNN achieved $0.0608$, significantly outperforming FF ($0.1579$) and the Transformer ($0.1811$), with reductions of $61.5\%$ and $66.4\%$, respectively. However, on the unseen test set, the Transformer achieved the best generalization performance, with a test MAE of $0.1675$, improving upon the RNN ($0.1857$) and FF ($0.2046$) by $9.8\%$ and $18.1\%$, respectively. These results suggest that while the RNN may fit the training data more closely, the Transformer demonstrates superior generalization to novel input distributions.

\noindent\textbf{Refinement Model (P2):}  
As shown in Figure~\ref{fig:performance-p2}, the Feedforward model exhibited the most consistent and generalizable performance in the second stage. It achieved the lowest validation MAE of $0.0715$ and a test MAE of $0.0660$, outperforming the RNN and Transformer by $19.2\%$ and $47.5\%$, respectively, in terms of test MAE. Although the RNN attained the lowest training loss, this did not translate into better generalization, with a test MAE of $0.0816$. The Transformer, once again, demonstrated the highest variability, yielding a substantially higher test MAE of $0.1258$.

\begin{figure}[t]
    \centering
    \includegraphics[width=0.8\linewidth]{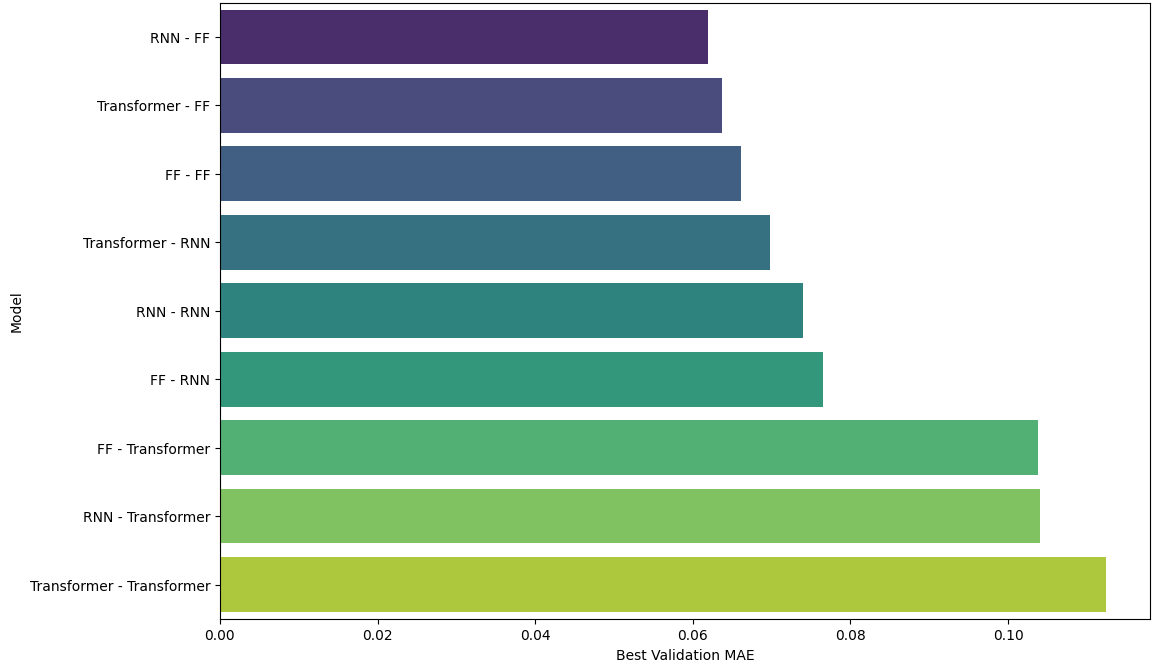}
    \caption{Combined validation results of P1-P2 model
pairs.}
    \label{fig:performance-p1-p2}
\end{figure}

\noindent\textbf{End-to-End Performance:}  
To evaluate the effectiveness of the two-phase scheduling in GOGH, we tested all pairwise combinations of P1 and P2 models. As shown in Figure~\ref{fig:performance-p1-p2}, the RNN--Feed Forward pipeline achieved the best overall performance, with a validation MAE of $0.06196$ and a validation loss of $0.01022$. Notably, this configuration outperformed the next best model, Transformer--Feed Forward, by approximately $2.8\%$ in MAE and $8.7\%$ in loss, highlighting its superior predictive capability. These findings indicate that capturing temporal patterns in the initial phase via an RNN, followed by deterministic refinement through a Feed Forward network, leads to more precise scheduling predictions. Overall, our results underscore that strategic model pairing within GOGH substantially affects final performance. The RNN--Feed Forward combination demonstrates the most favorable balance between accuracy and generalization, making it a promising candidate for deployment in real-world dynamic and heterogeneous computing environments.

\noindent\textbf{Discussion:}  
The system demonstrates a coherent and intuitive behavior in predicting job throughput across varying accelerator configurations. When evaluating unseen job combinations, the estimator consistently provides throughput estimates that closely reflect the actual observed performance trends. For example, jobs with similar model architectures or resource demands tend to be mapped to accelerators with comparable predicted throughput values, showing the system’s ability to capture structural similarities in job profiles. Furthermore, the estimator’s behavior under job perturbation—such as modifying batch sizes—remains stable. The qualitative pattern of predictions aligns well with expected performance degradation or improvement. For instance, increasing the batch size for a compute-intensive model typically leads to lower predicted throughput, which was corroborated by the empirical results. Another notable qualitative behavior is the system’s sensitivity to co-scheduling effects. When jobs with conflicting resource usage patterns are scheduled together, the estimator often anticipates contention, and the predicted throughput reflects this degradation. This mirrors the real-world dynamics observed during execution. Overall, these qualitative observations suggest that the estimator captures key structural properties of the workload and device interaction, providing interpretable and practically useful predictions even without relying on hand-crafted scheduling rules.

\section{Related Works}
In this section we review recent research that address the management of heterogeneous hardware through scheduling strategies, resource management techniques, and system architectures.

\noindent\textbf{Heterogeneity-Aware Scheduling.}
Narayanan et al. introduced Gavel~\cite{narayanan2020heterogeneity}, a scheduler that accounts for the performance variability of different accelerators like GPUs and TPUs across various deep learning models. By incorporating an "effective throughput" metric, Gavel optimizes job allocation to heterogeneous resources, improving overall system efficiency and reducing job completion times. Qiao et al. proposed Pollux~\cite{qiao2021pollux}, which dynamically adjusts resource allocation and training configurations based on real-time job performance. This co-adaptive approach enhances both system throughput and fairness among jobs, particularly in environments with diverse hardware capabilities.

\noindent\textbf{Resource Sharing and Utilization.}
Wu et al. developed TGS~\cite{wu2023transparent}, a system that enables transparent GPU sharing in containerized cloud environments. Operating at the OS level, TGS allows multiple deep learning jobs to share GPUs without significant performance degradation, thereby increasing GPU utilization and reducing energy waste. Zhao et al. presented HiveD~\cite{zhao2020hived}, a framework that ensures fair and efficient GPU cluster sharing among multiple tenants. By introducing the concept of Virtual Clusters and a hierarchical resource allocation strategy, HiveD maintains GPU affinity and prevents resource contention, leading to better utilization and energy efficiency.

\noindent\textbf{Serverless and Elastic Training Platforms.}
Gu et al. introduced ElasticFlow~\cite{gu2023elasticflow}, a serverless platform for distributed deep learning training. ElasticFlow allows users to specify training jobs without detailing resource requirements, automatically allocating resources to meet job deadlines. This elasticity leads to improved resource utilization and energy savings by adapting to the workload's needs.

\noindent\textbf{Optimizing Large Language Model Serving.}
Lin et al. proposed Parrot~\cite{lin2024parrot}, a system designed to enhance the serving of applications based on large language models (LLMs). By introducing "Semantic Variables," Parrot exposes application-level information to the serving system, enabling optimizations that reduce redundant computations and improve energy efficiency during inference.

\section{Conclusion}
In this paper, we addressed the challenge of managing GPUs in a heterogeneous deep learning cluster by proposing GOGH, a correlation-guided orchestration framework. GOGH leverages both inter-job and inter-GPU correlations to estimate the throughput of unseen jobs on various GPU types. Initial throughput estimates are refined using limited runtime observations on a subset of GPUs, enabling more accurate predictions with minimal overhead. We formulated an integer linear program (ILP) that utilizes these estimates to optimize GPU allocation, aiming to reduce power consumption while satisfying minimum throughput requirements for all jobs. To model and learn the underlying correlations, we employed multiple neural architectures, including feed-forward networks, recurrent neural networks (RNNs), and Transformers. Experimental results on a publicly available dataset demonstrate that GOGH achieves high estimation accuracy, with throughput prediction errors as low as $5\%$. Future directions include designing approximation algorithms to solve the ILP more efficiently and exploring reinforcement learning techniques to exploit correlation structures with even lower computational cost.
\bibliographystyle{unsrt}
\bibliography{references}

\end{document}